\documentclass[10pt,conference]{IEEEtran}
\IEEEoverridecommandlockouts

\usepackage{mathtools}
\usepackage{amsmath,amssymb}
\usepackage{paralist}
\usepackage{hyperref}
\usepackage{cleveref}
\usepackage{balance}
\usepackage{algorithm}
\usepackage{algorithmic}
\usepackage{tabularx}
\usepackage{multirow}
\usepackage{tikz}
\usepackage{booktabs}
\usepackage[citestyle=numeric,bibstyle=numeric,maxcitenames=1,mincitenames=1,natbib=true,doi=false,url=false,isbn=false]{biblatex}
\makeatletter
\newcommand{\linebreakand}{%
  \end{@IEEEauthorhalign}
  \hfill\mbox{}\par
  \mbox{}\hfill\begin{@IEEEauthorhalign}
}
\makeatother

\begin{document}

\title{An experience in automatically extracting CAPAs from code repositories}

\author{\IEEEauthorblockN{1\textsuperscript{st} Yegor Bugayenko}
\IEEEauthorblockA{\textit{Huawei}\\
Moscow, Russia \\
yegor256@huawei.com}
\and
\IEEEauthorblockN{2\textsuperscript{nd} Imre Delgado}
\IEEEauthorblockA{\textit{Innopolis University}\\
Innopolis, Russia \\
i.delgado@innopolis.university}
\and
\IEEEauthorblockN{3\textsuperscript{rd} Firas Jolha}
\IEEEauthorblockA{\textit{Innopolis University}\\
Innopolis, Russia \\
f.jolha@innopolis.university}
\and
\IEEEauthorblockN{4\textsuperscript{th} Zamira Kholmatova}
\IEEEauthorblockA{\textit{Innopolis University}\\
Innopolis, Russia \\
z.kholmatova@innopolis.university}
\linebreakand
\IEEEauthorblockN{5\textsuperscript{th} Artem Kruglov}
\IEEEauthorblockA{\textit{Innopolis University}\\
Innopolis, Russia \\
a.kruglov@innopolis.ru}
\and
\IEEEauthorblockN{6\textsuperscript{th} Witold Pedrycz}
\IEEEauthorblockA{\textit{University of Alberta}\\
Edmonton, Canada \\
wpedrycz@ualberta.ca}
\and
\IEEEauthorblockN{7\textsuperscript{th} Giancarlo Succi}
\IEEEauthorblockA{\textit{University of Bologna}\\
Bologna, Italy \\
g.succi@unibo.it}
\and
\IEEEauthorblockN{7\textsuperscript{th} Xavier Vasquez}
\IEEEauthorblockA{\textit{Innopolis University}\\
Innopolis, Russia \\
x.vasquez@innopolis.university}
}

\maketitle

\begin{abstract}
TOM (stands for Theoretically Objective Measurements of Software Development Projects) is a set of services that are in charge of helping developers or teams in the process of identifying anomilies within their software development process, and providing a list of preventive or corrective actions (aka CAPAS) that positively impact the process. and in this way to improve the quality of the final product and its development process.

In order to get help from TOM, it is as simple as adding our bot (@0capa) to the list of collaborators in your repository, and with this our bot will automatically take care of obtaining different metrics from your repository, in order to suggest actions to take into account to that in your future updates the identified anomalies are not repeated.

This paper presents the underlying research on this idea.  
\end{abstract}


\section{Introduction}
\label{S:Introducion}
We have been struggling with understanding the evolution of software projects for several years, in particular it is difficult to understand reasons and means of changes in code. We focus on this paper in how we can interpret information present in code repositories to determine the intended corrective and preventive maintenance actions performed by developers on the code.

The work is intended primarily to large organizations, featuring tens of thousands of developers or more, like Huawei, Microsoft, etc., and, indeed the overall world community of open source software development, which is actually the empirical tested for this study.

Software engineering is difficult and that despite now several decades of research, still most of the projects fail: according to The Standish Group less than 30\% of the projects can be considered successful, a percentage that has remained practically constant in the previous five years. Moreover, for the large projects the success rate is just above 10\%~\citep{CHAOS2015}. All these percentages should horrify project managers and key stakeholders of software development enterprises.

The proposed solutions appears to be monotonic: better process, better management, and more recently, more agility. To many it appears that it is only a matter of intensity and of higher discipline in adhering to protocols, and even reading more recent articles~\citep{Miranda22} we do not see a really change of perspective. Even if a plethora of alternative development processes have appeared, it appears that the intrinsic combinatorial complexity of all the factors determining how approach the wicked problems during software development process are completely ignored~\citep{Rittel1973}.

In a sense this is quite surprising, as from the ’70s there was a large
discussion of using expert systems in management~\citep{Eom1993} and the matter was already discussed by \citet{Ramsey1989}, when they suggested to use AI, and, specifically, expert systems to manage the complexity
of software projects. This idea went forward till the early 2000 with
various favours and contributions~\citep{Vasudevan}, a focus on effort estimation~\citep{Finnie},
and also using case based reasoning~\citep{GANESAN2000} and neural networks~\citep{Jun2001}. Their idea suffered from the relative inability of expert systems to be accepted by the general industry~\citep{Duchessi1995} and by the difficulty to formalize the problems of software management in terms of explicit knowledge.

The problem was then tackled from a different perspective by the work
of \citet{Robbins1998} and the one by \citet{Succi2000}; in these cases
instead of a predefined set of expert rules, the idea was to provide an
advice at runtime based on features that were being emerged by developing the project. This approach was strongly influenced by the emerging of the Agile approach, which centered the process in the developer~\citep{BecAnd04extreme} with the acknowledgment that software is the result of the creative act of the human mind. 

The scientific background and justification provide a basis to perform a significant move forward in the discipline, by elaborating mechanisms from data mining~\citep{Park2020} and machine learning~\citep{Kratsch2020} approaches that could retrieve the artifacts being produced by teams throughout the software development process, parse them, and compare them with previously developed ones, thereby identifying areas for improvements and suggesting best practices. Thus, the main research question to be answered in this research is: how measurable properties (metrics) of project digital artifacts can be attributed to suggestions of corrective and preventive actions (CAPAs), in order to have a positive effect.

The contribution of this research is the software agent and the model which provide specific recommendations in form of CAPAs based on the current state of a software development project.

The paper is structured as following. In \Cref{S:RelatedWorks} we observe the researches conducted in the area. \Cref{S:Methodology} presents the idea of the research at a high-level perspective, which further disclosed in \Cref{S:Pattern}, \Cref{S:CAPAs} and \Cref{S:Validation}. The description of the software that implements the developed model is given in \Cref{S:Implementation}. The conclusions on the work done are given in \Cref{S:Conclusion}.

\section{Related Work}\label{S:RelatedWorks}

A number of researchers in the software engineering community, intending to prevent project failures, proposed a variety of solutions aimed at improving both management~\citep{taherdoost2018lead} and process~\citep{moitra2005managing}. However, it is very difficult to choose a suitable management style or process because of the existence of a variety of factors that lead to software project failure, such as management style~\citep{cerpa2009did}, stakeholders involvement~\citep{charette2005software}, team~\citep{lehtinen2014perceived}, etc. Based on the critical analysis of 70 different recorded failed projects, \citet{verner2008factors} observed that for each of them, there is a wide spectrum of circumstantial factors (such as environmental and psychological ones) that could be considered the root causes of failures~\citep{lehtinen2014perceived}. It, therefore, seems that there is a strong need for objective and universal assessments of software projects, which could be achieved through the usage of metrics. The possibility of providing ``genuinely improved management decision support systems'' based on the adoption of specific software metrics has already been established~\citep{fenton1999software, el2001prediction, bhandari2018measuring}. 

Other researchers, such as~\citep{martinez2015artificial}, identified several management areas where artificial intelligent technologies or techniques could be applied. These include:
\begin{inparaenum}\label{E:managementAreas}
    \item estimation of project success;
    \item identification of critical success factors;
    \item relatedness to project budget;
    \item connection to projects schedule;
    \item project planning;
    \item relatedness to risk identification.
\end{inparaenum}

Yet, there is no item on this list that suggests actions that could be taken to prevent failures. However, \citet{zage1995avoiding} attempted to specify the relationship between software metrics and corrective and preventive actions.

The preliminary work we done in this subject area is related to the formation of an objective list of possible CAPAs. For this purpose we performed the analysis of the most popular project management frameworks, guides and standards to the software engineering and IT project management:
\begin{inparaenum}
    \item Control Objectives for Information and related Technology (COBIT),
    \item Information Technology Infrastructure Library (ITIL),
    \item Capability Maturity Model Integration (CMMI),
    \item The Guide to the Software Engineering Body of Knowledge (SWEBOK), and
    \item The Open Group Architecture Framework (TOGAF).
\end{inparaenum}
Based on expert evaluation of the applicability of these information sources, we chose CMMI for extraction of CAPAs relevant to the software development process. The resulted set of 766 derived CAPAs is available to our readers in the \href{https://github.com/AKruglov89/CAPAs/blob/main/CAPAs\%20-\%20CMMI\%20ver\%202.0.csv}{GitHub repository} for further analysis.

\section{Research workflow}
\label{S:Methodology}
The development of the model for theoretically objective measurements (TOM) a number of consecutive tasks must be solved. Thus, the general algorithm of the model can be represented as the following steps:
\begin{asparaenum}[Step 1.]
\item 
% In terms of code descriptors at a method level, a repository can be uniquely described as a matrix $X \in \mathbb{R}^{m \times n}$, where $m$ is the number of methods and $n$ is the number of metrics. The task at the first step is to define a set of metrics $M = (x_1, x_2, \dots , x_k),\ k<n$ that will describe the ``most'' of the variance in the code. 
A repository can be uniquely described by a multidimensional time series of specific metrics $T \in \mathbb{R}^{m \times n}$, where $n$ is the number of commits in the repository (or days of project life) and 
$m$ is the number of metrics. The task at the first step is to identify a minimum set $X = \{x_1,x_2,\dots,x_m\}$ of $m$ metrics that can comprehensively enough describe the status of the software project. 

\item Given a set of $m$ univariate time series of selected metrics $T_A = \{T^{(1)}, T^{(2)}, \dots, T^{(m)}\}$, we are looking for set of patterns $P = \{T_{i,j}^{(1)}, \dots, T_{i,j}^{(n)}\}$ whereby by $T_{i,j}$ is observed in $k$ time series belonging to $T_A$ and $i<j$. The pattern $T_{i,j}$ should appear not less than $a$ times and also not more than $b$ times in each of the $m$ univariate series. 

Once we identified such sets $\sum_{i=0}^N P_i$ for N repositories, we can look for similar patterns across repositories using Z-Normalized Euclidean Distance. 

The outcome of this step is a catalogue of patterns $\hat P= \{\hat p_1, \hat p_2, \dots, \hat p_l \}, l \le k$ repetitive across a number of repositories. 

\item After detecting a pattern we need to find the corrective action associated with it. In software repositories, any corrective action $c$ takes the form of a pull request. By extracting and analyzing the $m$ metrics of these $\hat p$ pull requests following the time of the given anomaly we can assign to every row $x_{i} \in X \in \mathbb{R}^{\hat p \times m}$ its respective corrective or preventive action (CAPA) $c_{k}$, leading to groups of pull requests according to their respective CAPAs.
Moreover, we can use the data of the relevant issue --- its open and close times --- to define more precisely the time range for the search. 

The outcome of this step should be the set of tuples $PC = \{\{\hat p_1,c_1\},\{\hat p_2,c_2\},\dots,\{\hat p_n,c_n\}\}$ for $n \le l$ which represent pairs of pattern $\hat p$ and related CAPA $c$.

% Going further, we can clarify the textual description of the CAPA by matching the predefined label of the cluster (e.g. ``add linter'' or ``refactoring'') with the results of the SLR or CMMI analysis (see \Cref{S:RelatedWorks}) using one of the NLP techniques. Then, the outcome of this step will be the set $\{\{a_1,\hat c_1\},\{a_2,\hat c_2\},\dots,\{a_n,\hat c_n\}\}$, where $\hat c$ is extended textual description of the CAPA $c$.

\item To understand whether the obtained relation between pattern and CAPA is not due to the randomness is the analysis of statistical distribution for each pair pattern-CAPA.

The resulted outcome of the whole methodology is a set $\sum_{i=0}^N \{\hat p_i,c_i\}$ with statistically significant pairwise relationship between elements. 
\end{asparaenum}

\section{Pattern Recognition}
\label{S:Pattern}

% Our objective was to create a pipeline, which takes numeric metrics collected from a source repository as an input, and suggests one of CAPAs to the project team as an output. The development, training and testing of the different modules of this solution is based on the post-mortem analysis of software development projects. Therefore, we avoid possible bias in subjective evaluation done by experts, and based our assumptions on historical data only.

% The algorithm is the following:

% \begin{asparaenum}[Step 1:]
By analysing the GitHub API we identified three metrics which could be collected for each commit and describe the ``most'' of the variance in the code: 
\begin{inparaenum}
\item total lines added,
\item total lines deleted, and
\item total lines changed.
\end{inparaenum}

We visually observed similarly shaped and repetitive figures in the diagrams of these metrics. We called them ``patterns'' and developed a method for automatically detecting them. The method has a few configuration parameters:
\begin{inparaenum}[a)]
\item the minimum percentage of repositories where a pattern must be seen,
\item the minimum and the maximum length of the pattern in data points,
\item the minimum and the maximum amount of times a pattern must be seen in a repository,
and
\item the minimum required Z-Normalized Euclidean Distance as a characteristic of similarity between all occurrences of a pattern.
\end{inparaenum}

The overall proposed algorithm for finding patterns of interest is presented in Algorithm \ref{alg:main}. The algorithm for pattern retrieval is available \href{https://github.com/Gci04/tom-pattern-recognition}{here}. 

\begin{algorithm}
\caption{Matrix Profile Approach}
\label{alg:main}
\renewcommand{\algorithmicrequire}{\textbf{Input:}}
\renewcommand{\algorithmicensure}{\textbf{Output:}}
\begin{algorithmic}[1]
    \REQUIRE $\gamma$ - set of time series , $p$ - maximum number of repositories to contain consensus pattern, $z$ - minimum number of times a consensus pattern is seen in each time series
	\ENSURE $\textbf{R}$ - set of consensus subsequences  % $T_{a,b}$ 
	\STATE $\textbf{R} \gets \emptyset$
	\FORALL{$m \in [d,e]$}
	    \STATE $k \gets 0$
        \STATE $T_{a,m} \gets$ consensus subsequence of length $m$
        % \STATE $Tweets_c \gets $All Tweets in class $c$
        \FORALL{$T \in \gamma$}
            \STATE $tmatches \gets $number of pattern matches $T_{a,m}$ in $T$
            \IF{$tmatches > z$}
                \STATE $k \gets k + 1$
            \ENDIF
        \ENDFOR
        \IF{$k < p$}
            \STATE add $T_{a,m}$ to $R$
        \ENDIF
    \ENDFOR
	\end{algorithmic}
\end{algorithm}

\begin{figure*}
    \includegraphics[width=\textwidth]{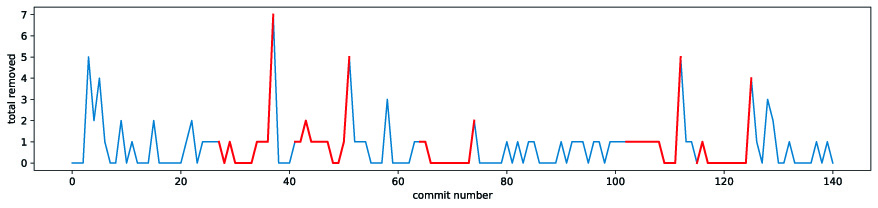}
    \caption{Sample repository represented by time series of lines removed and its corresponding patterns found using algorithm \ref{alg:main}. The red color highlights the patterns matching the consensus pattern}
    \label{fig:example3}
\end{figure*}

We tested the proposed solution on the \href{https://github.com/Gci04/tom-pattern-recognition/tree/main/Data}{dataset} of 1000 repositories and found 30 patterns of different types. The visualized example of a pattern is given in \Cref{fig:example3}. The patterns found are available \href{https://github.com/Gci04/tom-pattern-recognition/tree/main/Data}{here}. 

\section{CAPA Categorization}
\label{S:CAPAs}

It is a common practice in GitHub to make changes to repositories through pull requests (PR), which are reviewed by other contributors and then merged. We assumed that PRs are not homogeneous: most often they fix bugs and implement features, while sometimes they improve the entire code base. We assumed that the latter may be classified as CAPAs and created an automated ML-based method of such classification.
Using GitHub Search API and special keywords for each action, we found 10629 PRs (see \Cref{tab:train_test_first_experiment}). Using GitHub API, for each PR we collected 56 metrics and then, after a number of iterations, refined and reduced the list down to 27 metrics (see \Cref{tab:dataType}). 

\begin{table}
    \centering
    \caption{Size of train and test datasets for classification of pull requests into CAPAs and non-CAPAs.}
    \label{tab:train_test_first_experiment}
    \begin{tabularx}{\linewidth}{>{\raggedright\arraybackslash}Xrr}
    \toprule
    category & train & test \\
    \midrule 
 	 CAPAs 	 & 	 4743 	 & 	 1184	 \\
 	 non-CAPAs 	 & 	 5886 	 & 	 1474	 \\
    \midrule 
 	 Total   & 	 10629 	 & 	 2658	 \\
 	\bottomrule
    \end{tabularx}
\end{table}

\begin{table}
    \centering
    \caption{Data Type}
    \label{tab:dataType}
    \begin{tabularx}{\linewidth}{r>{\raggedright\arraybackslash}Xl}
        \toprule
        \# & metric & data type \\
        \midrule
        1 & number of comments & numeric \\
        2 & number of commits & numeric \\
        3 & number of files & numeric \\
        4 & number of issue comments & numeric \\
        5 & number of issue events & numeric \\
        6 & number of labels & numeric \\
        7 & number of review comments & numeric \\
        8 & number of review requests & numeric \\
        9 & number of reviewers & numeric \\
        10 & number of additions & numeric \\
        11 & closure date & timestamp \\
        12 & creation date & timestamp \\
        13 & number of deletions & numeric \\
        14 & locked state & boolean \\
        15 & merged state & boolean \\
        16 & merged date & timestamp \\
        17 & milestone status & boolean \\
        18 & milestone closure date & timestamp \\
        19 & number of milestone closed issues & numeric \\
        20 & milestone creation date & timestamp \\
        21 & milestone due on date & timestamp \\
        22 & milestone state & boolean \\
        23 & pull request number & numeric \\
        24 & pull request state & boolean \\
        25 & update date & timestamp \\
        26 & number of participants & numeric \\
        27 & number of file changes & numeric \\
        \bottomrule
    \end{tabularx}
\end{table}

We conducted series of experiments using three classification models. A Random Forest Classifier (RFC) with 500 estimators, an XGBoost Classifier (XBC) also 500 estimators, and a CatBoost Classifier (CBC) with a learning rate of 0.1, and 500 iteration steps. The experiments were performed on the dataset using a ratio of 80\% and 20\% for the train and test sets, respectively. 

First, we classified PRs into CAPAs and non-CAPAs (see \Cref{tab:cr_1}). Second, for the selected PRs we classified them to one of the possible actions (see \Cref{tab:cr_upsampled}). 

\begin{table}
    \centering
    \caption{Report of the first step of classification. Three models are compared: Random Forest Classifier (RFC), XGBoost Classifier (XBC), and CatBoost Classifier (CBC). Classes are labeled as: 1---CAPA category, 2---non-CAPA category.}
    \label{tab:cr_1}
    \tabcolsep=0.125cm
    \begin{tabularx}{\linewidth}{>{\raggedright\arraybackslash}X >{\raggedright\arraybackslash}X *{7}r}
    \toprule
         label & model & TP & TN & FP & FN & PRE & REC & F1 \\
% ******************************** 0 ********************************
    \midrule
    \multirow{3}{*}{1} 	 & 	 RFC 	 & 	 1174 	 & 	 1380 	 & 	 94 	 & 	 10 	 & 	 0.93 	 & 	 0.99 	 & 	 0.96	 \\
     	 & 	 XBC 	 & 	 1177 	 & 	 1378 	 & 	 96 	 & 	 7 	 & 	 0.92 	 & 	 0.99 	 & 	 0.96	 \\
     	 & 	 CBC 	 & 	 1169 	 & 	 1386 	 & 	 88 	 & 	 15 	 & 	 0.93 	 & 	 0.99 	 & 	 0.96	 \\

% ******************************** 1 ********************************
    \midrule
    \multirow{3}{*}{2} 	 & 	 RFC 	 & 	 1380 	 & 	 1174 	 & 	 10 	 & 	 94 	 & 	 0.99 	 & 	 0.94 	 & 	 0.96	 \\
     	 & 	 XBC 	 & 	 1378 	 & 	 1177 	 & 	 7 	 & 	 96 	 & 	 0.99 	 & 	 0.93 	 & 	 0.96	 \\
     	 & 	 CBC 	 & 	 1386 	 & 	 1169 	 & 	 15 	 & 	 88 	 & 	 0.99 	 & 	 0.94 	 & 	 0.96	 \\

 	 \bottomrule
    \end{tabularx}
\end{table}

\begin{table}
    \centering
    \caption{Report of the second step of classification. Each class is labeled as: 1---add linter, 2---coverage, 3---documentation, 4---functional requirements, 5---refactoring, 6---unstable build, 7---unused.}
    \label{tab:cr_upsampled}
    \tabcolsep=0.125cm
    \begin{tabularx}{\linewidth}{>{\raggedright\arraybackslash}X >{\raggedright\arraybackslash}X *{7}r}
    \toprule
     label & model & TP & TN & FP & FN & PRE & REC & F1 \\
% ******************************** 0 ********************************
    \midrule
    \multirow{3}{*}{1} 	 & 	 RFC 	 & 	 182 	 & 	 1187 	 & 	 10 	 & 	 21 	 & 	 \underline{0.95} 	 & 	 \underline{0.9} 	 & 	 \underline{0.92}	 \\
 	 & 	 XBC 	 & 	 176 	 & 	 1180 	 & 	 17 	 & 	 27 	 & 	 0.91 	 & 	 0.87 	 & 	 0.89	 \\
 	 & 	 CBC 	 & 	 169 	 & 	 1176 	 & 	 21 	 & 	 34 	 & 	 0.89 	 & 	 0.83 	 & 	 0.86	 \\
% ******************************** 1 ********************************
    \midrule
    \multirow{3}{*}{2} 	 & 	 RFC 	 & 	 184 	 & 	 1128 	 & 	 48 	 & 	 40 	 & 	 \underline{0.79} 	 & 	 \underline{0.82} 	 & 	 \underline{0.81}	 \\
 	 & 	 XBC 	 & 	 152 	 & 	 1129 	 & 	 47 	 & 	 72 	 & 	 0.76 	 & 	 0.68 	 & 	 0.72	 \\
 	 & 	 CBC 	 & 	 160 	 & 	 1114 	 & 	 62 	 & 	 64 	 & 	 0.72 	 & 	 0.71 	 & 	 0.72	 \\
% ******************************** 2 ********************************
    \midrule
    \multirow{3}{*}{3} 	 & 	 RFC 	 & 	 150 	 & 	 1183 	 & 	 39 	 & 	 28 	 & 	 \underline{0.79} 	 & 	 \underline{0.84} 	 & 	 \underline{0.82}	 \\
 	 & 	 XBC 	 & 	 125 	 & 	 1152 	 & 	 70 	 & 	 53 	 & 	 0.64 	 & 	 0.7 	 & 	 0.67	 \\
 	 & 	 CBC 	 & 	 117 	 & 	 1164 	 & 	 58 	 & 	 61 	 & 	 0.67 	 & 	 0.66 	 & 	 0.66	 \\
% ******************************** 3 ********************************
    \midrule
    \multirow{3}{*}{4} 	 & 	 RFC 	 & 	 200 	 & 	 1180 	 & 	 5 	 & 	 15 	 & 	 \underline{0.98} 	 & 	 0.93 	 & 	 \underline{0.95}	 \\
 	 & 	 XBC 	 & 	 202 	 & 	 1176 	 & 	 9 	 & 	 13 	 & 	 0.96 	 & 	 \underline{0.94} 	 & 	 \underline{0.95}	 \\
 	 & 	 CBC 	 & 	 194 	 & 	 1172 	 & 	 13 	 & 	 21 	 & 	 0.94 	 & 	 0.9 	 & 	 0.92	 \\
% ******************************** 4 ********************************
    \midrule
    \multirow{3}{*}{5} 	 & 	 RFC 	 & 	 169 	 & 	 1187 	 & 	 30 	 & 	 14 	 & 	 \underline{0.85} 	 & 	 \underline{0.92} 	 & 	 \underline{0.88}	 \\
 	 & 	 XBC 	 & 	 160 	 & 	 1154 	 & 	 63 	 & 	 23 	 & 	 0.72 	 & 	 0.87 	 & 	 0.79	 \\
 	 & 	 CBC 	 & 	 160 	 & 	 1154 	 & 	 63 	 & 	 23 	 & 	 0.72 	 & 	 0.87 	 & 	 0.79	 \\
% ******************************** 5 ********************************
    \midrule
    \multirow{3}{*}{6} 	 & 	 RFC 	 & 	 197 	 & 	 1174 	 & 	 15 	 & 	 14 	 & 	 \underline{0.93} 	 & 	 \underline{0.93} 	 & 	 \underline{0.93}	 \\
 	 & 	 XBC 	 & 	 180 	 & 	 1164 	 & 	 25 	 & 	 31 	 & 	 0.88 	 & 	 0.85 	 & 	 0.87	 \\
 	 & 	 CBC 	 & 	 178 	 & 	 1159 	 & 	 30 	 & 	 33 	 & 	 0.86 	 & 	 0.84 	 & 	 0.85	 \\
% ******************************** 6 ********************************
    \midrule
    \multirow{3}{*}{7} 	 & 	 RFC 	 & 	 160 	 & 	 1203 	 & 	 11 	 & 	 26 	 & 	 \underline{0.94} 	 & 	 \underline{0.86} 	 & 	 \underline{0.9}	 \\
 	 & 	 XBC 	 & 	 148 	 & 	 1188 	 & 	 26 	 & 	 38 	 & 	 0.85 	 & 	 0.8 	 & 	 0.82	 \\
 	 & 	 CBC 	 & 	 147 	 & 	 1186 	 & 	 28 	 & 	 39 	 & 	 0.84 	 & 	 0.79 	 & 	 0.81	 \\
 	 \bottomrule
    \end{tabularx}
\end{table}

The average demonstrated accuracy is 83\%.
The model is available \href{https://github.com/alimre/Split/blob/main/Model/Model\%20Description.md}{here}. The data used for training and testing is \href{https://github.com/alimre/Split/blob/main/Patterns\%20and\%20CAPAs/Detected\%20CAPAs.md}{here}.

\section{Validation of the Results}
\label{S:Validation}
For each instance of identified patterns, we scanned the corresponding repository for a pull request during the time period when the pattern occurred plus one month. This decision is dictated by the fact that the potential corrective action is not performed immediately after the associated pattern occurs, but after some time. At the end, we identified \href{https://drive.google.com/file/d/1h5lHwRYeU1Zss0-hnlNrbTFZtdD2IMBO/view}{217 pull requests} which then were categorized as specific CAPAs.

Our hypothesis is that there is a causality between some patterns and CAPAs. To prove it we need to test the statistical significance of the dependency between sets of detected patterns and CAPAs occurred nearby. All cases where the PR identified as a particular CAPA was created between the beginning of a particular pattern and one month after its end are shown in the \Cref{tab:CAPAs}

\begin{table}
    \centering
    \caption{Number of CAPAs that occurred in close proximity to the identified patterns}
    \label{tab:CAPAs}
    \tabcolsep=0.125cm
    \begin{tabularx}{\linewidth}{l *{8}{>{\raggedleft\arraybackslash}X}}
    \toprule
    \multirow{2}{*}{Pattern type}  & \multicolumn{8}{c}{CAPA} \\
    & 0 & 1 & 2 & 3 & 4 & 5 & 6 & Total \\ 
    \midrule
    Pattern 5&6&2&5&4&0&3&0&20 \\
    Pattern 6&2&2&1&1&0&0&0&6 \\
    Pattern 7&2&0&6&2&0&3&0&13 \\
    Pattern 8&4&3&4&0&0&0&1&12 \\
    Pattern 9&3&6&7&1&1&2&1&21 \\
    Pattern 10&3&6&5&0&3&2&1&20 \\
    Pattern 11&0&9&9&0&0&11&2&31 \\
    Pattern 12&8&7&19&0&0&6&3&43 \\
    Pattern 13&8&6&3&1&1&4&0&23 \\
    Pattern 14&1&8&10&1&1&4&3&28 \\
\midrule
Total&37&49&69&10&6&35&11&217 \\
\bottomrule
\end{tabularx}
\end{table}

We ran Chi-squared test of independence to check whether there is a relationship between patterns and CAPAs. Our null hypothesis was that there is no relationship between patterns and CAPA, With the obtained p-value 0.0085 we can reject the null hypothesis with a level of significance alpha 0.05a one-sample t-test to check the dependency between patterns and CAPAs. The result of experiment showed the presence of dependence with p-value 0.0085.

% For each type of pattern we empirically determined the rule for the CAPAs to be considered as potentially relevant to it as at least five time of occurrences. 

For each type of pattern we empirically determined the rule for the CAPAs to be considered as potentially relevant to it as at least five time of occurrences. Then for the filtered tuples pattern-CAPA we performed a pairwise comparison to understand whether the specific CAPA has a significantly higher probability of belonging to the given pattern. \Cref{tab:CAPAs_2test} shows the results of this experimentinvestigation. For example, the first row represents the investigation of the following null hypothesis: the mean value of probability that CAPA 0 belongs to Pattern 5 is equal to the mean value of probability that CAPA 2 belongs to Pattern 5. While testing the hypothesis, we got p-value equal to 0.136. If we consider the level of significance 0.05, the null hypotheses cannot be rejected.

% \begin{table}
%     \centering
%     \caption{Average values of probabilities of belonging patterns to CAPAs}
%     \label{tab:CAPAs_2test}
%     \begin{tabularx}{\linewidth}{>{\raggedright\arraybackslash}X *{4}{>{\raggedleft\arraybackslash}X}}
%     \toprule
%     Pattern type & CAPA0 & CAPA1 & CAPA2 & CAPA5 \\
%     \midrule
%     Pattern 5 & 0.7 & - & 0.54 & - \\
%     Pattern 9 & - & 0.48 & 0.46 & - \\
%     Pattern 10 & - & 0.54 & 0.48 & - \\
%     Pattern 11 & - & 0.63 & 0.43 & 0.45 \\
%     Pattern 12 & 0.94 & 0.67 & 0.48 & 0.48 \\
%     Pattern 13 & 0.73 & 0.55 & - & - \\
%     Pattern 14 & - & 0.7 & 0.54 & - \\
% \bottomrule
% \end{tabularx}
% \end{table}

\begin{table}
    \centering
    \caption{Pairwise comparison of the CAPAs with 2-sample t-test}
    \label{tab:CAPAs_2test}
    \begin{tabularx}{\linewidth}{l *{5}{>{\raggedleft\arraybackslash}X}}
    \toprule
    Pattern type & (I) Spot & (J) Spot & Mean value I & Mean value J & p-value \\
    \midrule
    Pattern 5 & CAPA 0 & CAPA 2 & 0.7 & 0.54 & 0.136  \\
    \midrule
    Pattern 9 & CAPA 1 & CAPA 2 & 0.48 & 0.46 & 0.806  \\
    \midrule
    Pattern 10 & CAPA 1 & CAPA 2 & 0.54 & 0.48 & 0.512 \\
    \midrule
    \multirow{3}{*}{Pattern 11} & CAPA 1 & CAPA 2 & 0.63 & 0.43 & 0.003  \\
     & CAPA 1 & CAPA 5 & 0.63 & 0.45 & 0.006 \\
     & CAPA 2 & CAPA 5 & 0.43 & 0.45 & 0.71 \\
    \midrule
    \multirow{6}{*}{Pattern 12} & CAPA 0 & CAPA 1 & 0.94 & 0.67 & 0.001  \\
     & CAPA 0 & CAPA 2 & 0.94 & 0.48 & $2.6\cdot e^{-8}$ \\
     & CAPA 0 & CAPA 5 & 0.94 & 0.48 & $3.5\cdot e^{-5}$ \\
     & CAPA 1 & CAPA 2 & 0.67 & 0.48 & 0.007 \\
     & CAPA 1 & CAPA 5 & 0.67 & 0.48 & 0.021 \\
     & CAPA 2 & CAPA 5 & 0.48 & 0.48 & 0.93 \\
    \midrule
    Pattern 13 & CAPA 0 & CAPA 1 & 0.73 & 0.55 & 0.112  \\
    \midrule
    Pattern 14 & CAPA 1 & CAPA 2 & 0.7 & 0.54 & 0.004 \\
\bottomrule
\end{tabularx}
\end{table}

As a result of the analysis, we showed a statistically significant relationship between certain patterns and CAPAs. By choosing a acceptable level of significance, we can determine the final set of patterns-CAPA. Thus, for the significance level 0.15, the output of the whole experiment is a dictionary of the following tuples pattern-CAPA: 
\begin{equation*}
    \begin{gathered}
    (Pattern\ 5 \rightarrow CAPA\ 0), \\
    (Pattern\ 11 \rightarrow  CAPA\ 1), \\
    (Pattern\ 12 \rightarrow  CAPA\ 0), \\
    (Pattern\ 13 \rightarrow  CAPA\ 0), \\
    (Pattern\ 14 \rightarrow  CAPA\ 1). 
    \end{gathered}
\end{equation*}

\section{Implementation}
\label{S:Implementation}
In order to automate of the data collection and processing process, a system was developed that is described below.

\begin{figure}
    \centering
    \includegraphics[width=\columnwidth]{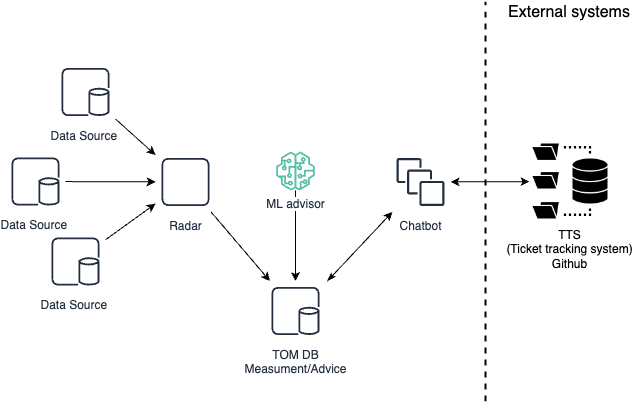}
    \caption{System Architecture - High level view}
    \label{fig:arch1}
\end{figure}

Given the needs of the system, a system was proposed that is composed of 3 modules mainly (see \Cref{fig:arch1}), where the first of them is in charge of integrating and monitoring the different sources of information, carrying out an initial implementation to integrate GitHub, this particular component will be called ``TOM-Radar''.

The next of the developed components is the one in charge of processing the collected data, relying on the model that was described in \Cref{S:Methodology}, an ML module was developed, which is capable of taking the different metrics collected by the radar and produce a set anomalies.

Finally, to complete the system, a module was developed that is in charge of notifying the teams about the different anomalies found, in a way that CAPAS can propose to solve the incidents detected.

\begin{figure*}
    \centering
    \includegraphics[width=.8\linewidth]{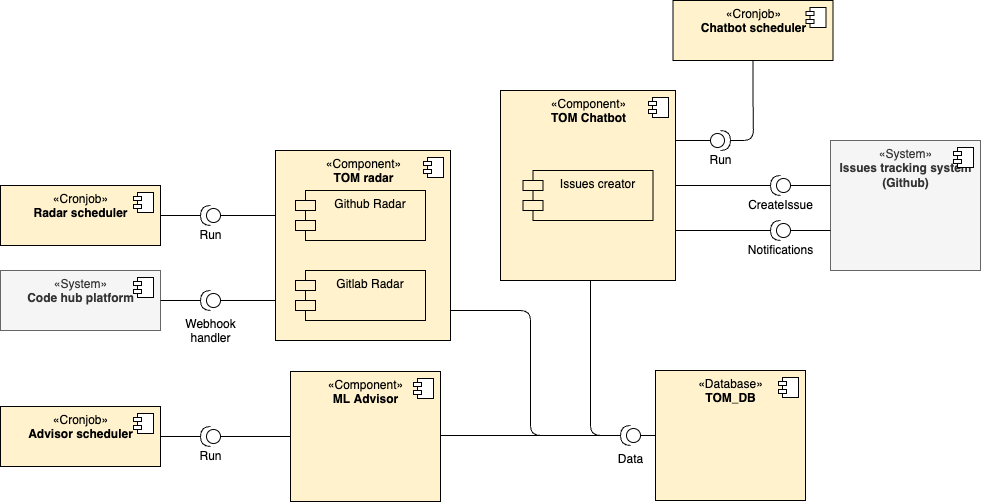}
    \caption{System Architecture - components view}
    \label{fig:arch2}
\end{figure*}

Now, diving deeper in the modules’ implementation, we can realize that both 3 modules (see \Cref{fig:arch2}) have been designed following a similar pattern, since for all of them we want to meet similar non-functional requirements. So that, taking into consideration that one of the main drivers is the flexibility of the system to be able to be integrated into multiple data sources, implement different processing models and, in the same way, be able to integrate with different ticket tracking systems.

\begin{figure}
    \centering
    \includegraphics[width=.7\columnwidth]{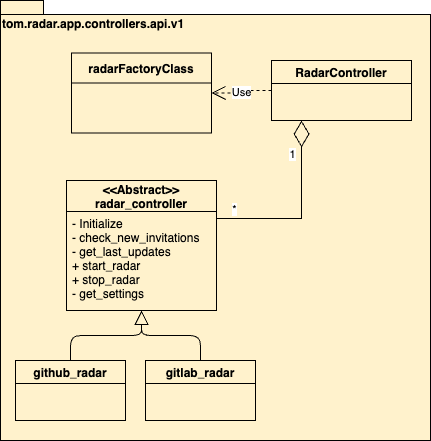}
    \caption{System Architecture - radar implementation}
    \label{fig:arch3}
\end{figure}

Regarding the implementation of the TOM-Radar module, in the picture \Cref{fig:arch3}, we can see that it is a quite simple but quite functional design, where given the need to be able to generate different instances of the radars, an instance factory has been implemented that depending on the load of the system can dispatch them as needed, then looking at the details of the radar itself, we see that we have defined the methods necessary to complete the data collection cycle, which must be implemented in each type of radar that you want to incorporate.

Where we find methods to initialize the radar, in which basically the radar configurations are loaded and the state is updated to control the concurrence of the different instances, in addition to this, we see another method that is responsible for monitoring if the system has received permissions to monitor new repositories or projects, to be able to add them to the processing queue and a third method that takes care of data collection per se.

And as a control measure, two methods were added to be able to start and stop the radar activity, in this way the steps described above are executed periodically and unattended.
Seeing this level of detail, it is worth mentioning that both the ML advisor and the chat-bot have a similar design, with an instance factory and a clearly defined template for integration with new external systems.

The developed software is available \href{https://github.com/cqfn/0capa}{here}.

\section{Conclusion}
\label{S:Conclusion}
In this research we have demonstrated the idea of managing the software development process through theoretically objective measurement. We performed the post-mortem analysis of more than 1000 repositories to identify the typical characteristic situations in development process, which we called patterns. Also, we analyzed more than 10000 pull requests to create a model for their classification into specific corrective and preventive actions. After it, we identified the tuples pattern-CAPA which have statistically significant dependency. 

This research could be considered as a spike solution to what we can call automatic monitoring and management of software development projects. It demonstrates the overall concept, however, to make the solution effective we need to expand the CAPA set, add pattern retrieval from multivariate time series, and increase the number of analyzed repositories to tens of thousands. 

\balance
\printbibliography

@article{Rittel1973,
  doi = {10.1007/bf01405730},
  url = {https://doi.org/10.1007/bf01405730},
  year = {1973},
  month = jun,
  publisher = {Springer Science and Business Media {LLC}},
  volume = {4},
  number = {2},
  pages = {155--169},
  author = {Horst W. J. Rittel and Melvin M. Webber},
  title = {Dilemmas in a general theory of planning},
  journal = {Policy Sciences}
}

@article{Eom1993,
  doi = {10.1016/0377-2217(93)90309-b},
  url = {https://doi.org/10.1016/0377-2217(93)90309-b},
  year = {1993},
  month = jul,
  publisher = {Elsevier {BV}},
  volume = {68},
  number = {2},
  pages = {278--290},
  author = {Sean B. Eom and Sang M. Lee and Ahmed Ayaz},
  title = {Expert systems applications development research in business: A selected bibliography (1975{\textendash}1989)},
  journal = {European Journal of Operational Research}
}

@article{Ramsey1989,
  doi = {10.1109/32.24728},
  url = {https://doi.org/10.1109/32.24728},
  year = {1989},
  month = jun,
  publisher = {Institute of Electrical and Electronics Engineers ({IEEE})},
  volume = {15},
  number = {6},
  pages = {747--759},
  author = {C.L. Ramsey and V.R. Basili},
  title = {An evaluation of expert systems for software engineering management},
  journal = {{IEEE} Transactions on Software Engineering}
}

@inproceedings{Vasudevan,
  doi = {10.1109/tai.1994.346435},
  url = {https://doi.org/10.1109/tai.1994.346435},
  publisher = {{IEEE} Comput. Soc. Press},
  author = {C. Vasudevan},
  title = {An experience-based approach to software project management},
  booktitle = {Proceedings Sixth International Conference on Tools with Artificial Intelligence. {TAI} 94},
  year = {1994}
}

@inproceedings{Finnie,
  doi = {10.1109/seep.1996.534020},
  url = {https://doi.org/10.1109/seep.1996.534020},
  publisher = {{IEEE} Comput. Soc. Press},
  author = {G.R. Finnie and G.E. Wittig},
  title = {{AI} tools for software development effort estimation},
  booktitle = {Proceedings 1996 International Conference Software Engineering: Education and Practice},
  year = {1996}
}

@article{GANESAN2000,
  doi = {10.1142/s0218194000000092},
  url = {https://doi.org/10.1142/s0218194000000092},
  year = {2000},
  month = apr,
  publisher = {World Scientific Pub Co Pte Lt},
  volume = {10},
  number = {02},
  pages = {139--152},
  author = {K. Ganesan and Taghi M. Khoshgoftaar and Edward B. Allen},
  title = {Case-based Software Quality Prediction},
  journal = {International Journal of Software Engineering and Knowledge Engineering}
}

@article{Jun2001,
  doi = {10.1016/s0957-4174(01)00021-5},
  url = {https://doi.org/10.1016/s0957-4174(01)00021-5},
  year = {2001},
  month = jul,
  publisher = {Elsevier {BV}},
  volume = {21},
  number = {1},
  pages = {1--14},
  author = {Eung Sup Jun and Jae Kyu Lee},
  title = {Quasi-optimal case-selective neural network model for software effort estimation},
  journal = {Expert Systems with Applications}
}

@article{Duchessi1995,
  doi = {10.1016/0957-4174(94)00056-2},
  url = {https://doi.org/10.1016/0957-4174(94)00056-2},
  year = {1995},
  month = jan,
  publisher = {Elsevier {BV}},
  volume = {9},
  number = {2},
  pages = {123--133},
  author = {Peter Duchessi and Robert M. O{\textquotesingle}Keefe},
  title = {Understanding expert systems success and failure},
  journal = {Expert Systems with Applications}
}

@inproceedings{Robbins1998,
  doi = {10.1145/268389.268416},
  url = {https://doi.org/10.1145/268389.268416},
  year = {1998},
  publisher = {{ACM} Press},
  author = {Jason E. Robbins and David M. Hilbert and David F. Redmiles},
  title = {Software architecture critics in Argo},
  booktitle = {Proceedings of the 3rd international conference on Intelligent user interfaces  - {IUI} {\textquotesingle}98}
}

@inproceedings{Succi2000,
  doi = {10.1145/337180.337641},
  url = {https://doi.org/10.1145/337180.337641},
  year = {2000},
  publisher = {{ACM} Press},
  author = {Giancarlo Succi and Jason Yip and Eric Liu and Witold Pedrycz},
  title = {Holmes},
  booktitle = {Proceedings of the 22nd international conference on Software engineering  - {ICSE} {\textquotesingle}00}
}

@book{BecAnd04extreme,
  address = {Boston},
  author = {Beck, Kent and Andres, Cynthia},
  description = {Extreme Programming Explained, second edition},
  isbn = {0321278658},
  publisher = {Addison-Wesley Professional},
  title = {Extreme Programming Explained: Embrace Change (2nd Edition)},
  url = {http://portal.acm.org/citation.cfm?id=1076267},
  year = 2004
}

@incollection{Park2020,
  doi = {10.1007/978-3-030-66498-5_16},
  url = {https://doi.org/10.1007/978-3-030-66498-5_16},
  year = {2020},
  publisher = {Springer International Publishing},
  pages = {206--218},
  author = {Gyunam Park and Wil M. P. van der Aalst},
  title = {A General Framework for Action-Oriented Process Mining},
  booktitle = {Business Process Management Workshops}
}

@article{Kratsch2020,
  doi = {10.1007/s12599-020-00645-0},
  url = {https://doi.org/10.1007/s12599-020-00645-0},
  year = {2020},
  month = apr,
  publisher = {Springer Science and Business Media {LLC}},
  volume = {63},
  number = {3},
  pages = {261--276},
  author = {Wolfgang Kratsch and Jonas Manderscheid and Maximilian R\"{o}glinger and Johannes Seyfried},
  title = {Machine Learning in Business Process Monitoring: A Comparison of Deep Learning and Classical Approaches Used for Outcome Prediction},
  journal = {Business {\&} Information Systems Engineering}
}

@misc{CHAOS2015,
  author = {{The Standish Group}},
  title = {Chaos Report},
  year = 2015,
  note = {\url{https://www.standishgroup.com/sample_research_files/CHAOSReport2015-Final.pdf}
       -- last visited $08^{th}$ of May, 2022}
}

@misc{Miranda22,
  author = {Dana Miranda and Adam Hardy},
  title = {Core Project Management Methodologies---And How To Choose The One For You},
  year = 2022,
  note = {\url{https://www.forbes.com/advisor/business/project-management-methodologies/}
       -- last visited $08^{th}$ of May, 2022}
}

@article{taherdoost2018lead,
  author  = {Hamed Taherdoost},
  title   = {{How to Lead to Sustainable and Successful IT Project Management? Propose 5Ps Guideline}},
  journal = {Propose 5Ps Guideline (August 1, 2018)},
  year    = {2018},
  eprint = {https://dx.doi.org/10.2139/ssrn.3224225}
}

@inproceedings{bhandari2018measuring,
  author       = {Guru Prasad Bhandari and Ratneshwer Gupta},
  booktitle    = {Proceedings of the 2018 5th IEEE Uttar Pradesh Section International Conference on Electrical, Electronics and Computer Engineering (UPCON)},
  title        = {{Measuring the Fault Predictability of Software Using Deep Learning Techniques with Software Metrics}},
  organization = {IEEE},
  pages        = {1--6},
  year         = {2018},  doi={10.1109/UPCON.2018.8597154}
}

@article{cerpa2009did,
  author    = {Narciso Cerpa and June M. Verner},
  title     = {{Why Did Your Project Fail?}},
  issn      = {0001-0782},
  number    = {12},
  pages     = {130--134},
  volume    = {52},
  journal   = {Communications of the ACM},
 publisher = {Association for Computing Machinery},
  year      = {2009},
  address = {New York, NY, USA},
  issue_date = {December 2009},
  url = {https://doi.org/10.1145/1610252.1610286},
  doi = {10.1145/1610252.1610286},
}

@article{charette2005software,
  author    = {Robert N. Charette},
  title     = {{Why Software Fails [Software Failure]}},
  issn      = {0018-9235},
  number    = {9},
  pages     = {42--49},
  volume    = {42},
  journal   = {IEEE Spectrum},
  publisher = {IEEE},
  year      = {2005},
  doi={10.1109/MSPEC.2005.1502528}
}

@article{el2001prediction,
  author    = {Khaled El Emam and Walcelio Melo and Javam C. Machado},
  title     = {{The Prediction of Faulty Classes Using Object-oriented Design Metrics}},
  issn      = {0164-1212},
  number    = {1},
  pages     = {63--75},
  volume    = {56},
  journal   = {Journal of Systems and Software},
  publisher = {Elsevier Science Inc.},
  address = {USA},
  year      = {2001},
  issue_date = {Feb 1,2001},
  url = {https://doi.org/10.1016/S0164-1212(00)00086-8},
  doi = {10.1016/S0164-1212(00)00086-8},
  numpages = {13}
}

@article{fenton1999software,
  author    = {Norman E. Fenton and Martin Neil},
  title     = {{Software Metrics: Successes, Failures and New Directions}},
  issn      = {0164-1212},
  number    = {2-3},
  pages     = {149--157},
  numpages = {9},
  volume    = {47},
  journal   = {Journal of Systems and Software},
  publisher = {Elsevier Science Inc.},
  address = {USA},
  year      = {1999},
  issue_date = {July 1, 1999},
  url = {https://doi.org/10.1016/S0164-1212(99)00035-7},
  doi = {10.1016/S0164-1212(99)00035-7},
}

@article{lehtinen2014perceived,
  author    = {Timo O.A. Lehtinen and Mika V. Mäntylä and Jari Vanhanen and Juha Itkonen and Casper Lassenius},
  title     = {{Perceived Causes of Software Project Failures--an Analysis of Their Relationships}},
  issn      = {0950-5849},
  number    = {6},
  pages     = {623--643},
  volume    = {56},
  journal   = {Information and Software Technology},
  publisher = {Elsevier},
  year      = {2014},
  doi = {https://doi.org/10.1016/j.infsof.2014.01.015},
  url = {https://www.sciencedirect.com/science/article/pii/S0950584914000263},
}

@article{martinez2015artificial,
  author    = {Daniel Magaña Martínez and Juan Carlos Fernandez-Rodriguez},
  title     = {{Artificial Intelligence Applied to Project Success: A Literature Review}},
  issn      = {1989-1660},
  number    = {5},
  pages     = {77--84},
  volume    = {3},
  journal   = {International Journal of Interactive Multimedia and Artificial Intelligence},
  year      = {2015},
}

@inbook{moitra2005managing,
  author    = {Deependra Moitra},
  booktitle = {Software Process Modeling},
  title     = {{Managing Organizational Change for Software Process Improvement}},
  pages     = {163--185},
  publisher={Springer US},
  address={Boston, MA},
  year      = {2005},
  editor={Silvia T. Acu{\~{n}}a and Natalia Juristo},
  doi={10.1007/0-387-24262-7_7},
  url={https://doi.org/10.1007/0-387-24262-7_7}
}

@inproceedings{verner2008factors,
  author       = {June Verner and Jennifer Sampson and Narciso Cerpa},
  booktitle    = {Proceedings of the 2008 second international conference on research challenges in information science},
  title        = {{What Factors Lead to Software Project Failure?}},
  organization = {IEEE},
  pages        = {71--80},
  year         = {2008},
  doi={10.1109/RCIS.2008.4632095}
}

@article{zage1995avoiding,
  author    = {Wayne M. Zage and Dolores M. Zage and Cathy Wilburn},
  title     = {{Avoiding Metric Monsters: A Design Metrics Approach}},
  issn      = {1022-7091},
  number    = {1},
  pages     = {43--55},
  volume    = {1},
  journal   = {Annals of Software Engineering},
  publisher = {Springer},
  year      = {1995},
  doi = {10.1007/BF02249045},
  url = {https://doi.org/10.1007/BF02249045}
}

% \bibliographystyle{IEEEtran}
% \bibliography{references.bib}

\end{document}